\documentclass[twocolumn, showkeys,preprintnumbers,%
amsmath,amssymb,prc]{revtex4}
\usepackage{graphicx}% Include figure files
\usepackage{bm}% bold math
\begin{document}
\title{Visualisation of an entangled channel spin-1 system}

\author{Swarnamala\,\,  Sirsi}
\author{Veena Adiga}
\altaffiliation {[Also at :\,St.Joseph's College(autonomous), Bengaluru -27,India]}
\email{vadiga11@gmail.com} 
\affiliation{Department of Physics, Yuvaraja's College \\
University of Mysore, Mysore-05,\,India}

\date{\today}

\begin{abstract}
Co-variance matrix formalism gives powerful entanglement criteria for continuous as well as finite dimensional systems. We use this formalism to study a mixed channel spin-1 system which is well known in nuclear reactions. A spin-j state can be visualized as being made up of 2j spinors which are represented by a constellation of 2j points on a Bloch sphere using Majorana construction. We extend this formalism to visualize an entangled mixed spin-1 system. 
\end{abstract}

\maketitle

Entanglement occurs as a result of quantum interference of states giving rise to non-classical correlations between spatially separated quantum systems. Understanding and manipulating dynamics of these quantum features are of great importance for both fundamental physics and new emerging quantum technologies. Entanglement is needed in several quantum information processing tasks such as quantum cryptography \cite{1}, quantum teleportation \cite{2}  and quantum computation \cite{3}. Characterisation of entangled states is of great importance to understand the underlying mathematical structure of a given state. Entanglement of a bipartite system in a pure state is unambiguous
and well defined. In contrast, mixed-state entanglement is relatively poorly understood. The purpose of this paper is to visualize the notion of entanglement in statistical assemblies of particles with spin-1. There are several criteria \cite{4} known to characterize entanglement like concurrence, entanglement of formation (EOF), positive partial transpose (PPT) etc. For a detailed discussion of this we refer
the reader to the recent articles \cite{4,5}. Here we employ co-variance matrix formalism  \cite{6,7} as the elements of co-variance matrix are closely  related to intrinsic quantum correlations that exist between the constituent spinors of a spin system. Such correlations  are shown to explain the physical origin of the squeezing behavior of the coupled spin-1 system \cite{8,9}. It can be noted that the spin squeezing inequalities generalize the concept of spin squeezing parameter and provide the necessary and sufficient condition for genuine two qubit entanglement of spin-1 system \cite{10}. Also an equivalence between the Peres-Horodecki criterion (PPT) and negativity of the covariance matrix  has been  established \cite {6} showing that the condition is both necessary and sufficient for two qubit entanglement.\\
   
The correlations can be classified as those that arise (i) due to the coupling of the two subsystems (ii) due to the projection of the  combined total density matrix  `$\rho_{c}$' onto the desired spin space. In this paper we have taken $\rho_{c}$  to be a direct product of the two subsystem density matrices $\rho(1)$  and $\rho(2)$  which are not  entangled and has no correlations of the first kind. However when we take the spin-1 projection of  $\rho_{c}$   , the correlation of second kind will appear. \\

Our motivation here is to study entanglement of a channel spin-1 system which can be realised employing polarised spin 1/2 beam and polarised spin 1/2 target. Such systems naturally arise in nuclear physics experiments like hadron scattering and reaction processes \cite{11,12,13,14,15}. This discussion is best done  by employing the language of density matrix which can be applied with equal ease to pure as well as mixed spin systems. We outline the statistical tensor formalism using the well known spherical tensor representation for the density matrix and obtain the expressions for co-variance matrix elements which represent  spin-spin correlations between the constituent spinors. Further we go on to show that spin-1 density matrix can be visualized using Majorana construction \cite{16}. \\

It is well known that a symmetric two qubit state in which the two qubits are completely symmetric under interchange is confined to 3-dimensional Hilbert-Schmidt space spanned by the eigen states of the total angular momentum  of the qubits $[ |j = 1,m\rangle; m = \pm 1,0 ]$. It has been shown that \cite{6} in the case of symmetric states the co-variance matrix defined through 
\begin{equation}
C_{ij} = [\langle\sigma_{1i}\sigma _{2j}\rangle - \langle\sigma _{1i}\rangle\langle\sigma _{2j}\rangle ] 
\end{equation}
where $\sigma _{1i} = \sigma _{i}\otimes I$\,, $\sigma _{2i} = I\otimes \sigma _{i}$\,\,\,( I is the $2\times 2$ identity matrix and  $\sigma _i$ are the standard Pauli spin matrices), is necessarily positive semi-definite for separable symmetric state. It is also shown  that negativity of the covariance matrix is  a necessary and sufficient condition for  entanglement in symmetric states. 

 The standard expression for the density matrix `$\rho$' for a spin j system is 
\begin{equation}
\rho = \frac {Tr(\rho)}{(2j+1)}\sum^{2j}_{k=0}\,\sum^{+k}_{q=-k}\,\, t^{k}_{q}\, \tau^{k^{\dagger}}_{q}
\end{equation}
where $\tau^{k}_{q}$ \, (with $\tau^{0}_{0} = I$ , the identity operator) are irreducible tensor operators of rank `k' in the 2j+1 dimension spin space with projection `q' along the axis of quantization in the real 3-dimensional 
space. The $\tau^{k}_{q}$ satisfy the orthogonality relations
\begin{equation}
Tr({\tau^{k^{\dagger}}_{q}\tau^{k^{'}}_{q^{'}}})= (2j+1)\,\delta_{kk^{'}} \delta_{qq^{'}}.
\end{equation}
Here the normalization has been chosen so as to be in agreement with Madison convention\cite{17}. The spherical tensor parameters $t^{k}_{q}$ which characterize the given system are the average expectation values given by 
\begin{equation}
t^{k}_{q} = \frac {Tr({\rho\,\tau^{k}_{q}})}{Tr\rho}.
\end{equation}
Since $\rho$ is Hermitian and $\tau^{k^{\dagger}}_{q} = (-1)^{q}\tau^{k}_{-q}$ ,  $t^{k}_{q}$ satisfy the condition 
\begin{equation}
t^{k^{*}}_{q} = (-1)^{q}\,t^{k}_{-q}
\end{equation}

Let us now consider the example of channel spin-1 system which plays an important role in nuclear reactions. A beam of nucleons colliding with a proton target provides such an example. If both the beam and the target are prepared to be in mixed states, then the corresponding density matrices are given by 
\begin{equation}
\rho(i) = \frac{1}{2}\,[\, I + \vec\sigma (i)\cdot\vec p(i)\,] = \frac {1}{2}\,\sum_{k,q}\,t^{k}_{q}(i)\,\tau^{k^{\dagger}}_{q}(i);\,\, \, i=1,2.
\end{equation}
where ${\vec p(i)}$ are the polarization vectors and $\vec\sigma(i)'$s are the Pauli spin matrices.

The combined density matrix is the direct product of the individual matrices.  
\begin {equation}
\rho_{c} = \rho(1) \otimes \rho(2)
\end{equation}
Explicitly for spin-1 system \cite{9},
\begin{eqnarray}
t^{1}_{q} = \left [\frac{\sqrt{6}}{3+\vec p(1)\cdot \vec p(2)}\right]\,(\vec p_{q}(1)+\vec p_{q}(2))\\
t^{2}_{q} = \left [\frac {2\sqrt{3}}{3+\vec p(1)\cdot \vec p(2)}\right]\,(\vec p(1)\otimes\vec p(2))^{2}_{q}.
\end{eqnarray} 
   
Let us consider the special Lakin frame (SLF) \cite{9} which is widely used in studying nuclear reactions as follows : Choose $\hat z_0$ to be along $\vec p(1)+\vec p(2)$. Since $\vec p(1)$, $\vec p(2)$ together define a plane in any general situation, we choose ${\hat x_0}$\, to be in this plane such that the azimuths of ${\vec p(1)}$ , $\vec p(2)$ with respect to  $\hat x_0$  are respectively 0 and ${\pi}$. The $\hat y_0$ axis is then chosen to be along ${\hat z_0\times \hat x_0}$. The frame so chosen is indeed the special Lakin frame (SLF)\, as it is clear from above equations (8) and (9) that $t^{1}_{\pm1} = 0$ and $t^{2}_{2} = t^{2}_{-2}$ .
In this frame, the cartesian components of polarization vectors are given by
\begin{eqnarray}
p_{x_{0}}(1) &=& \frac {p(1)p(2)sin2\theta}{|\vec p(1)+\vec p(2)|} = -p_{x_{0}}(2)\\
p_{y_{0}}(1) &=& p_{y_{0}}(2)=0                                                    \\
p_{z_{0}}(1) &=& \frac {p(1)^{2}+p(1)p(2)cos2\theta}{|\vec p(1)+\vec p(2)|}\,\,\,;\nonumber\\
p_{z_{0}}(2) &=& \frac {p(2)^{2}+p(1)p(2)cos2\theta}{|\vec p(1)+\vec p(2)|},
\end{eqnarray}
where $2\theta$ is the angle between $\vec p(1)$\, and \,\,$\vec p(2)$.

Choosing a simple case of $|\vec p(1)| = |\vec p(2)|= p $, we get $t^{2}_{\pm1} = 0 $ in SLF.
The  density matrix so obtained is compared with that of the symmetric state density matrix  given in  equation (17) of reference (6).  It is observed that the co-variance matrix  takes the canonical form and the diagonal elements are given by \\
\begin{eqnarray}
C_{x_{0}x_{0}} \, &=&{\frac {1}{3}-\frac {\sqrt{2}}{3}\,t^{2}_{0}+\frac {2}{\sqrt{3}}\,t^{2}_{2}} 
                  = {\frac {1-p^{2}(1+2sin^{2}\theta)}{(3+p^{2}cos2\theta)}}\\
C_{y_{0}y_{0}}\, &=& \frac {1}{3}-\frac {\sqrt{2}}{3}\,t^{2}_{0}-\frac {2}{\sqrt {3}}\,t^{2}_{2} 
                 = {\frac {1-p^{2}cos2\theta}{(3+p^{2}cos2\theta)}}\\
C_{z_{0}z_{0}} &=& \frac {1}{3}+\frac {2\sqrt{2}}{3}\,t^{2}_{0}-\left({\sqrt\frac{2}{3}}\,t^{1}_{0}\right)^{2}\nonumber
\\
&=& \frac {1+p^{2}(1+2cos^{2}\theta)}{(3+p^{2}cos2\theta)}-\left(\frac{4pcos{\theta}}{(3+p^{2}cos2\theta)}\right)^{2} 
\end{eqnarray}
\begin{figure}
\includegraphics[width=8.6cm]{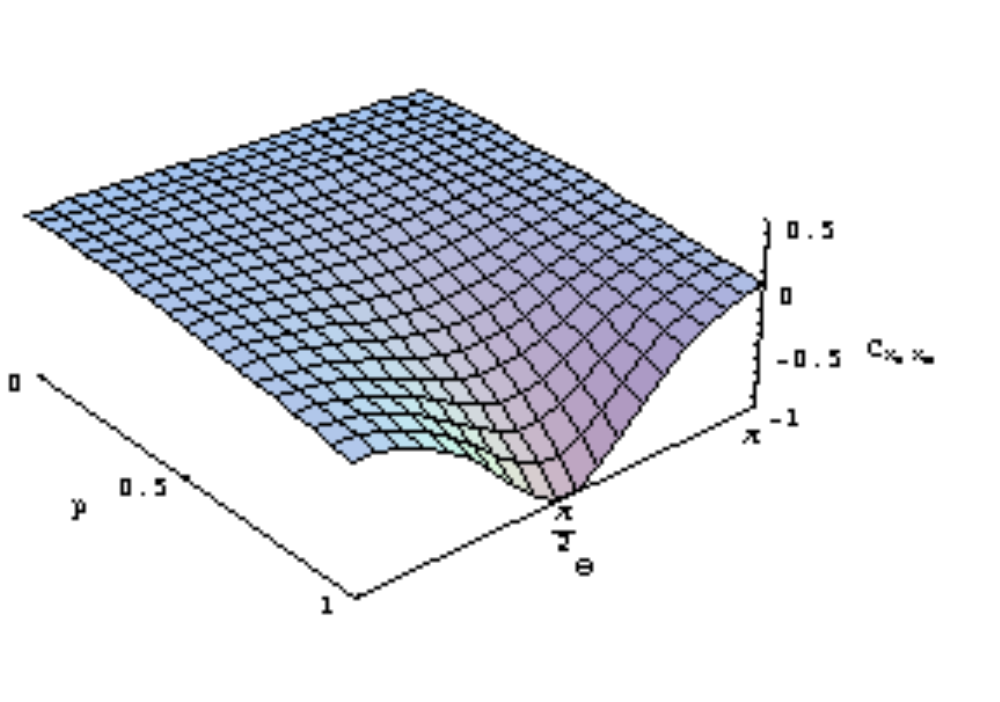}
\vspace{-5mm}
\caption{\label{fig:epsart} (color online) Variation of $C_{x_{0}x_{0}}$   as  function of  beam and target polarization p and $\theta$(rad). $C_{x_{0}x_{0}}< 0$ indicates entanglement.}
\includegraphics[width=8.6cm]{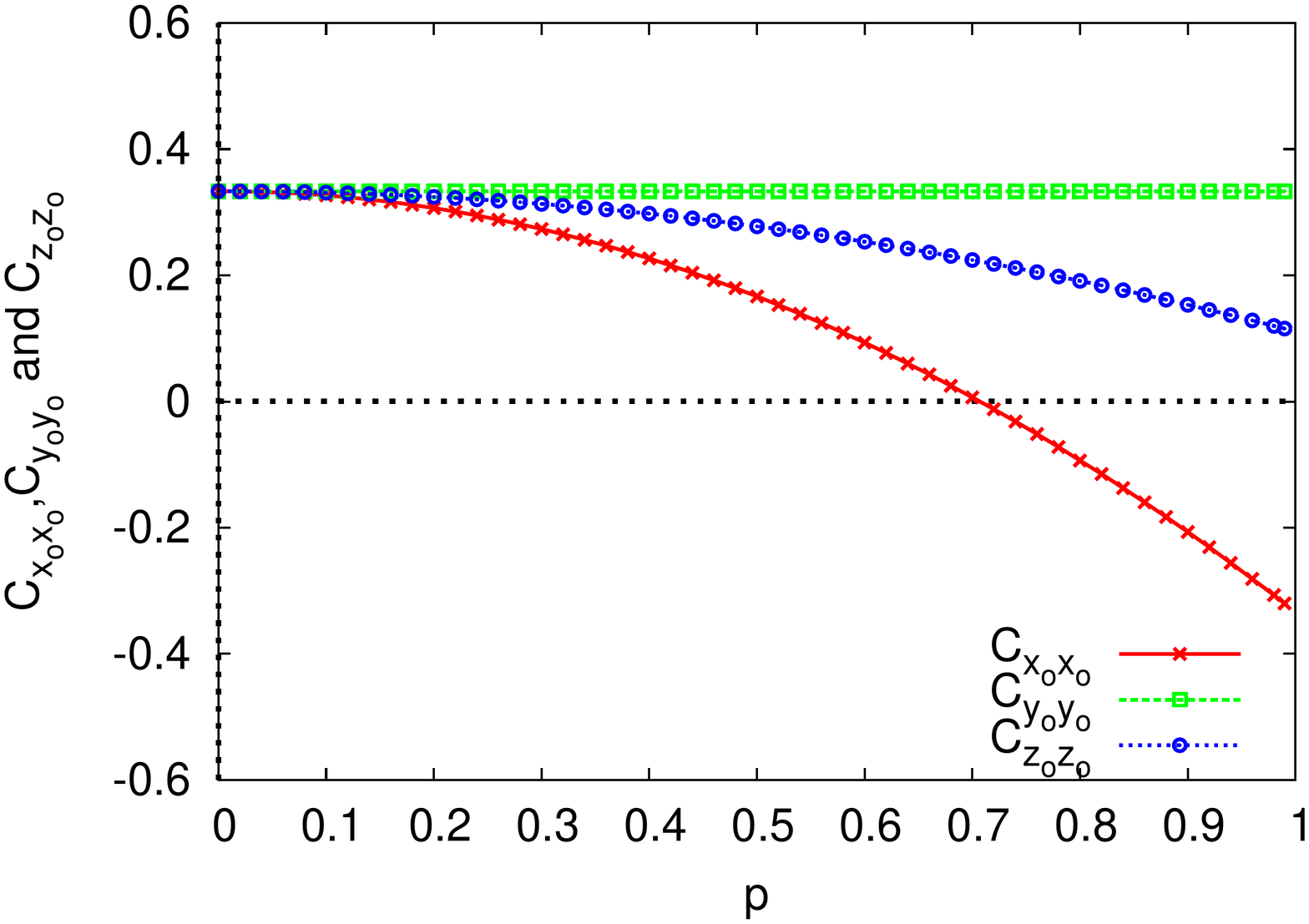}
\vspace{-2mm}
\caption{\label{fig:epsart} (color online) Variation of $C_{x_{0}x_{0}}$,\,$C_{y_{0}y_{0}}$ and $C_{z_{0}z_{0}}$  as a  function of beam and target polarization p for $\theta = \pi/4 $. $C_{x_{0}x_{0}}< 0$ indicates entanglement.}
\includegraphics[width=8.6cm]{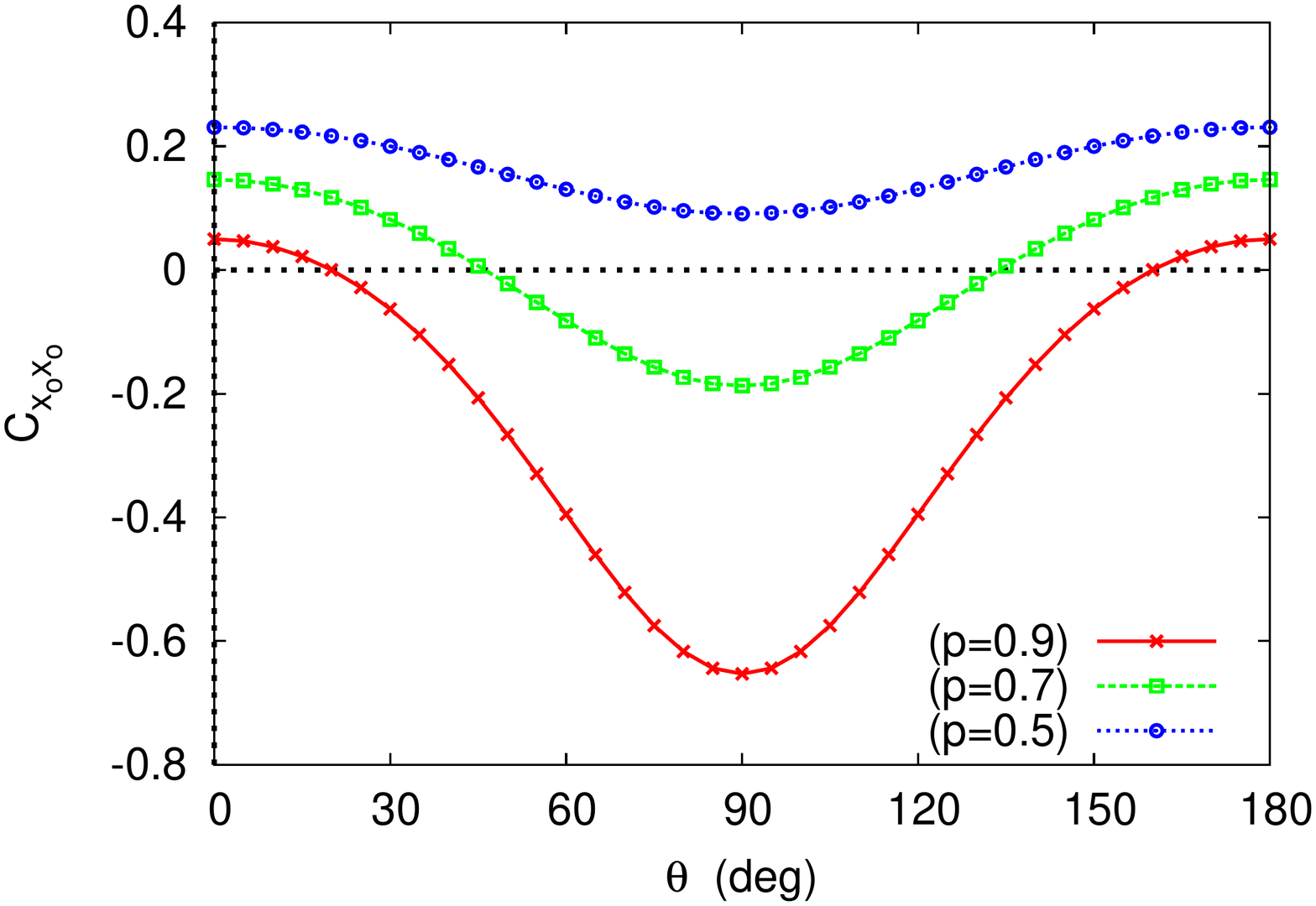}
\vspace{-2mm}
\caption{\label{fig:epsart} (color online) Variation of $C_{x_{0}x_{0}}$\, as a  function of angle $\theta$ (deg) for  beam and target polarization p = 0.5,\, p = 0.7 and p = 0.9. $C_{x_{0}x_{0}}< 0$ indicates entanglement.}
\end{figure}
\begin{figure}
\includegraphics[width=8.6 cm]{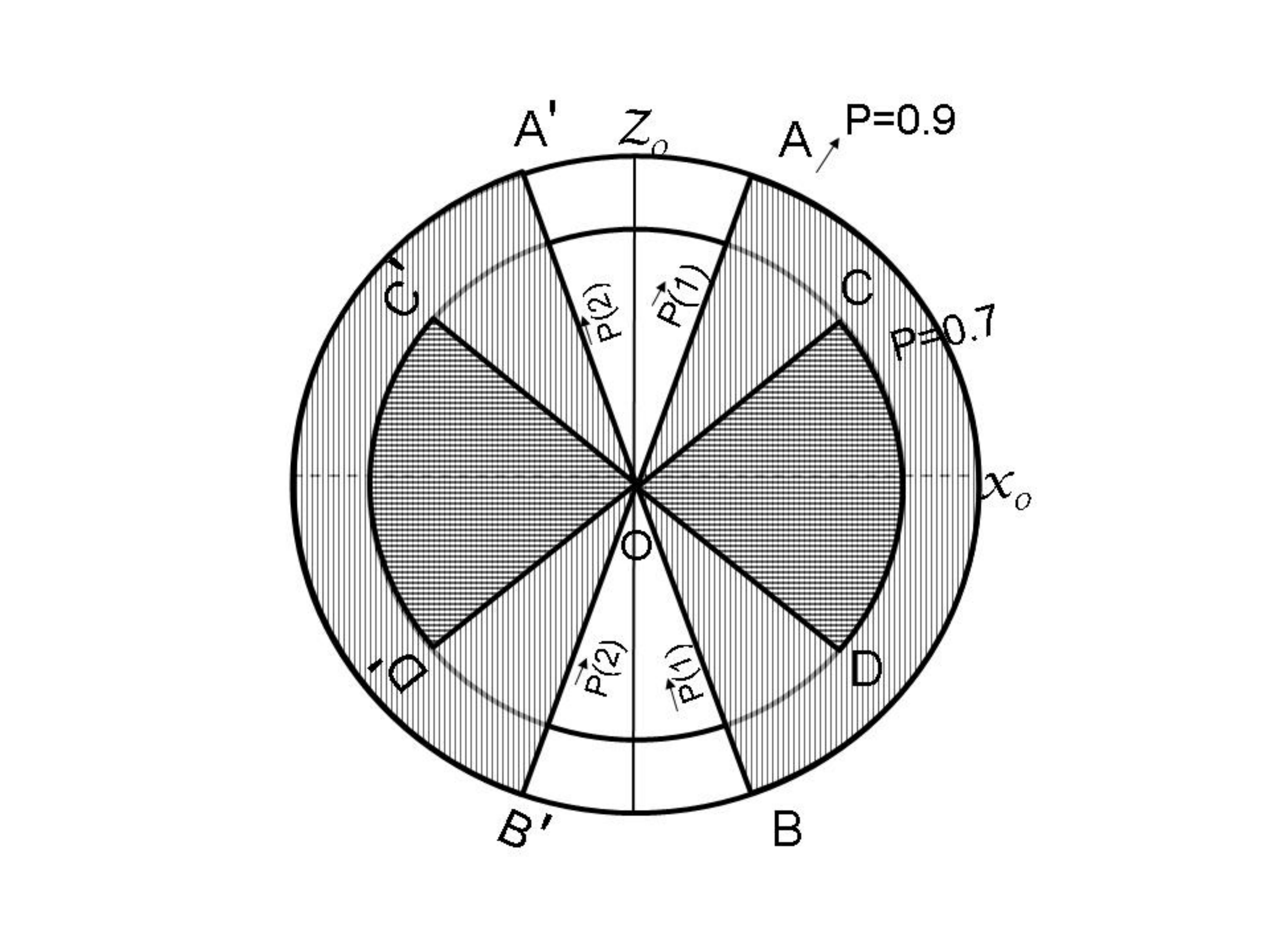}
\vspace{-2mm}
\caption{\label{fig:epsart} (color online) Range of $\theta$ for which the system is entangled for beam and target polarization p=0.9 (vertical lines) and p=0.7 (horizontal lines). Extreme lines of the entangled regions represent the polarisation vectors in SLF.} 
\end{figure}

Variation of $C_{x_{0}x_{0}}$  as function of $\theta$ (rad), beam and target polarization p is shown in fig(1). Variation of $C_{x_{0}x_{0}}$, $C_{y_{0}y_{0}}$ and $C_{z_{0}z_{0}}$  as  a function of p for $\theta = \pi/4$ is shown in fig(2).  Note that even if one of the diagonal elements is negative\,, the state is entangled . Also variation of $C_{x_{0}x_{0}}$ as  function of $\theta$ for  three values of p (p = 0.5, p = 0.7 and p = 0.9) is shown in fig(3). Range of $\theta$ for which the system is entangled is shown in fig(4). AOB, $A^{\prime}OB^{\prime}$ represent the region of entanglement for $\vec p(1)$\,, \,$\vec p(2)$ respectively in SLF when p = 0.9. Similarly COD, $C^{\prime}O D^{\prime}$ represent the region of entanglement when p = 0.7. It is seen that as p increases range of entanglement also increases. In particular, when p = 1, i.e., when the states are pure, the combined system will also be in a pure state and the spin-1 projection of this pure state will be in an entangled state except for $\theta$ = 0 and $\pi $ (equation (25) of reference (8)).

In order to visualize the entangled channel spin-1 system, let us discuss the geometric representation of the most general mixed spin-j density matrix. It has $(n^{2}-1)$ independent parameters  where $n=(2j+1)$ and is written as   
$\rho = \sum _{i=1}^{n} \lambda _{i} |\psi _{i}\rangle \langle \psi _{i}|$\,\,\,\,\,\,.
Here $\lambda _{i}'s$ are the distinct eigen values belonging to $|\psi _{i}\rangle$. (This can  be easily generalized to the degenerate case also).

If all of these $|\psi_{i}\rangle^{'} s$ are the eigen kets of $J_{z}$, then the number of independent parameters gets reduced to n+1. ((n-1) weights and $(\theta,\phi)$ of the quantization axis). Such a system is said to be oriented \cite{18}. In all other cases atleast one of the $|\psi_{i}\rangle ^{'}s$  is not an eigen ket of $J_{z}$. Such a system is said to be non-oriented spin system. In the most general non-oriented system none of the $|\psi_{i}\rangle^{'}s$ is an eigen ket of $J_{z}$. Such a system can be parameterized using Majorana construction \cite{16} according to which a spin-j state can be written as 
\begin{eqnarray}
|\psi ^{j}\rangle &=& \mathcal {N} \sum _{\mathcal{P}} \mathcal{\hat{P}}\{|\epsilon _{1}\epsilon _{2}---\epsilon _{2J}\rangle\} 
 \\
\mbox{where,} |\epsilon _{r}\rangle &=& \left(\begin{matrix}
cos(\alpha _{r}/2)\\
sin(\alpha _{r}/2)\, e^{i\beta _{r}}\\ 
\end{matrix}\right) \nonumber
\end{eqnarray}
are the 2j spinors constituting the spin-j state. $\hat {\mathcal{P}}$ is a set of $N!$ permutations and $\mathcal {N}$ is the normalization factor. Since $|\psi^{j}\rangle$ can be expanded in terms of angular momentum basis vectors as $|\psi^{j}\rangle = \sum^{+j}_{m=-j}\mathcal C^{j}_{m}\, |jm\rangle$\,\,\,\,, the Majorana polynomial can be written as  
\begin{equation}
P(Z) = \sum _{k=0}^{2j} (-1)^{k} \sqrt {\left (\begin{matrix}k\\ 2j \end{matrix}\right )}\,\,\, d_{k} Z^{k}
\end{equation}
where k=j+m and $d_{j+m} = \mathcal C^{j}_{m}$ are the complex co-efficients. Also ${\left (\begin{matrix}k\\ 2j \end{matrix}\right )}$ represents the binomial co-efficients and $ Z = tan (\alpha_{r}/2)\,e^{i\beta_{r}}$. These polynomials determine the orientation $(\alpha_{r},\beta_{r})$ of the constituent spinors \cite{19}, which can be visualized as a constellation of 2j points on the Bloch sphere. 

Considering the particular example of channel spin-1 system in SLF, the eigen values $\lambda _{1}$, $\lambda _{2}$, and $\lambda _{3}$ of the density matrix are found to be
\begin{eqnarray}
\lambda _{1,2}&=&\frac {1}{N}[1+p^{2}cos^{2}\theta \pm \sqrt {(p^4 sin^{4}\theta + 4p^{2}cos^{2}\theta)}]
\end{eqnarray}
\begin{eqnarray}
\lambda _{3}&=&\frac {1-p^{2}}{N},
\end{eqnarray}
where $ N = 3+p^{2}cos(2\theta)$.
The corresponding eigen vectors, $|\psi _{i}\rangle$ ; i = 1, 2 and $|\psi _{3}\rangle$ are
\begin{eqnarray}
|\psi _{i}\rangle &=& \frac{1}{N_{i}}
[p^{2}sin^{2}\theta|11\rangle \nonumber \\ &+& ((1+pcos\theta)^{2}-\lambda_{i}N)|1-1\rangle]  \\
|\psi _{3}\rangle &=& |10 \rangle.
\end{eqnarray}
where $N_{i}=(p^{4}sin^{4}\theta+((1+pcos\theta)^{2}-\lambda_{i}N)^{2})^{1/2}$.\\

Substituting for ${d_{j+m}}^{'}\, s$ in equation (17), $(\alpha,\beta)$ of the two spinors for the state $|\psi _{1}\rangle$ are given by $( 2\, tan^{-1}\sqrt{|x|},0)$ and $( 2\, tan^{-1}\sqrt{|x|},\pi)$ where $x = \displaystyle{ {((1+pcos\theta)^{2}-\lambda _{1}N})/{p^{2}sin^{2}\theta}}$. Observe that the spinors for the state  $|\psi _{1}\rangle$ are confined to $x_{o}-z_{o}$ plane. Similar construction for the state $|\psi_{2}\rangle$ indicates that the two spinors  are in the $y_{o}-z_{o}$ plane with  $(\alpha,\beta)$  given by $( 2\, tan^{-1}\sqrt{|y|} ,\pi/2)$and $(2\, tan^{-1}\sqrt{|y|} ,3\pi/2)$ where $y = \displaystyle{\ {((1+pcos\theta)^{2}-\lambda _{2}N})/{p^{2}sin^{2}\theta}}$. Since $|\psi_{3}\rangle$ is the  $|10\rangle$ state, it is constituted by up and down spinors.\\

In this paper we have studied the entanglement of a mixed spin-1 system as a function of polarisation of spin 1/2 beam ($\vec p(1)$) and spin 1/2 target ($\vec p(2)$) using co-variance matrix formalism. In the particular case of 
$|\vec p(1)| = |\vec p(2)|= p $, it is found that as p increases the range of $\theta$ for which the system is entangled also increases where $2\theta$ is the inclusive angle between $\vec p(1)$ and $\vec p(2)$. When the beam and the targets are in pure state, the resultant channel spin-1 system will also be in pure state and is entangled except for $\vec p(1)$ parallel and antiparallel to $\vec p(2)$. Further, using Majorana construction we have visualised channnel spin-1 system  as being made up of three sets of two spinors whose polar and azimuthal angles are functions of p and $\theta$. It is found that two of the spinorial sets are confined to $x_{o}-z_{o}$ plane and $y_{o}-z_{o}$ plane of a Bloch sphere respectively and the remaining two are given by up and down spinors with respect to $z_{o}$-axis. Thus any entangled state can be visualised in terms of constituent spinors.
Since none of the entanglement measures could capture the essence of mixed state entanglement, we believe that a deeper understanding of geometric characterisation of the Bloch sphere representation may lead to an intuitive understanding of entanglement. Our analysis can also be used to establish a quantitative relationship between spin-spin correlations and entanglement in higher spins. A detailed study of all these aspects using density matrix formalism is underway. 
\begin{acknowledgments}
The authors thank the referee for comments in the light of which the paper has been revised. One of us (VA) acknowledges with thanks the support provided by the University Grants Commission (UGC), INDIA for
the award of teacher fellowship through Faculty Development Programme (FDP).
\end{acknowledgments} 

\end{document}